\documentclass[12pt]{article}
\usepackage{url}
\usepackage{amsmath}
\usepackage{times}
\usepackage[english]{babel}
\usepackage{amsthm}
\usepackage{amssymb}
\usepackage[pdftex]{graphicx}
\usepackage{subfigure}
\usepackage[numbers,comma, sort&compress]{natbib}
\usepackage[bookmarks=true,colorlinks=true,linkcolor=blue,
urlcolor=blue,citecolor=blue]{hyperref}

\title{\Large\bf  Longitudinal excitations in triangular lattice antiferromagnets}
\author{Mohammad Merdan and Y Xian \\
\small School of Physics and Astronomy,\\
\small The University of Manchester, Manchester M13 9PL, UK}
\begin{document}
\maketitle
\begin{abstract}

We study the longitudinal excitations of quantum antiferromagnets on a triangular lattice by a recently proposed microscopic many-body approach based on magnon-density waves. We calculate the full longitudinal excitation spectra of the antiferromagnetic  Heisenberg model for a general spin quantum number in the isotropic limit. Similar to the square lattice model, we find that, at the center of the first hexagonal Brillouin zone $\Gamma(\mathbf q=0)$ and at the magnetic ordering wavevectors $\pm[\mathbf Q= (4\pi/3,0)]$, the excitation spectra become gapless in the thermodynamic limit, due to the slow, logarithmic divergence of the structure factor. However, these longitudinal modes on two-dimensional models may be considered as quasi-gapped, as any finite-size effect or small anisotropy will induce a large energy gap, when compared with the counterpart of the transverse spin-wave excitations. We also discuss a possible second longitudinal mode in the triangular lattice model due to the noncollinear nature of its magnetic order.
\end{abstract}

\newpage
\section {Introduction}
Anderson's spin-wave theory (SWT) provides a good description for the low temperature properties of many two-dimensional (2D) and three-dimensional (3D) quantum antiferromagnetic systems on a bipartite lattice \cite{Anderson}, including quantum corrections to the classical N\'{e}el ground state with two alternating sublattices and the doublet transverse spin-wave excitations. The quantum antiferromagnetic systems on a triangular lattice, however, are the prototypal system with the frustrated spin alignments in the classical picture and have been under intensive study over the last few decades for fundamentally different type of ground states \cite{ANDERSON1973,Fazek1974,Kalmeyer1987}. It is now firmly established by various methods \cite{ springerlink:10.1007/BFb0119592,Springer-Verlag.816.135}, including a SWT based one three-sublattices \cite{Huse1,Jolicoeur,Singh1,Miyake1992, Bernu,Azaria, Elstner1,Chubukov1994, Manuel1, Adolfo2000,Mezzacapo2010}, that the ground state of the antierromagnetic Heisenberg model has the long-ranged noncollinear order of the $120^{\circ}$ magnetic three-sublattice structure with three transverse, gapless spin-wave excitations.

Most isotropic antiferromagnets in one-dimension (1D) with low quantum spin numbers do not show N\'{e}el-like long-ranged order in the ground state due to the strong quantum fluctuations. The low-lying excitation states are also different from the 2D and 3D counterparts. In particular, by the exact solutions using the Bethe ansatz \cite{Bethe1931}, the low-lying excitation states of the 1D spin-1/2 Heisenberg model have been shown corresponding to the spin-1/2 object (spinons) where the spin-wave-like excited states are the triplet states of spinons which always appear in pairs \cite{14}, contrast to the doublet excitation states by SWT; the excitation states of spin-1 Heisenberg model for the linear chain, including the longitudinal one, have an excitation gap above the singlet ground state, first predicted by Haldane \cite{PhysRevLett.50.1153}. These theoretical predictions have later been confirmed in the antiferromagnetic compound KCuF$_3$ for spin-1/2 chains \cite{PhysRevLett.70.4003} and CsNiCl$_3$ for spin-1 chains \cite{PhysRevLett.56.371} by neutron-scattering experiments. One interesting remaining question is whether or not there exist longitudinal excitations in quantum antiferromagnetic systems with long-ranged classical order in low temperature, as such modes will represent the oscillations in the magnitude of the long-ranged order parameter. The answer is affirmative. There is now ample evidence of the longitudinal excitation states in various quasi-1d structures with the N\'eel-like long-ranged order at low temperature, including the hexagonal $ABX_3$-type antiferromagnets with both spin quantum number $s=1$ (CsNiCl${}_3$ and RbNiCl${}_3$)  \cite{Steiner1987,Tun1990} and $s=5/2$ (CsMnI${}_3$) \cite{PhysRevB.43.679,Kenzelmann2002} and the tetragonal structure of KCuF${}_3$ with $s=1/2$ \cite{lake2005}. More recently, a longitudinal mode was also observed in the dimerized antiferromagnetic compound TlCuCl${}_3$ under pressure with a long-ranged N\'eel order \cite{Ruegg2008}. To our knowledge, no observation of longitudinal modes in any 2D or quasi-2d antiferromagnets has been reported yet. Clearly, such longitudinal modes, which correspond to the oscillations in the magnitude of the magnetic order parameter, are beyond the usual SWT which predicts only the transverse spin-wave excitations, usually referred to as quasiparticle magnons in the antiferromagnetic systems. There are several theoretical investigations in these longitudinal modes using the field theory approach, such as a simplified version of Haldane's theory for the spin-1 systems \cite{Affleck1989,Affleck1992} or the sine-Gordon theory for the spin-1/2 systems \cite{Schulz1996,Essler1997}, and both treating the inter-chain couplings as perturbations. A phenomenological field theory approach focusing on the spin frustrations of the hexagonal lattice of the $ABX_3$-type antiferromagnetic systems has also been made \cite{Plumer1992}. We recently proposed a microscopic many-body theory based on the magnon-density waves for the longitudinal excitations of spin-$s$ quantum antiferromagnetic systems, using the original bipartite spin-lattice Hamiltonians \cite{Xian2006,Xian2007}. The basic physics in our analysis follows Feynmann's theory on the low-lying excited states of the helium-4 superfluid \cite{Feynman1954,Feynman1956}: the longitudinal excitation states in a quantum antiferromagnet with a N\'eel-like order are identified as the collective modes of the magnon-density waves, which represent the fluctuations in the long-range order and are supported by the interactions between magnons; these longitudinal excitation states are constructed by the magnon density operator $s^z$ in contrast to the transverse spin-flip operator $s^\pm$ of the magnon states \cite{yang2011}. Our numerical results \cite{yang2011} for the energy gap values at the magnetic wave vector are in good agreement with the experiments for the energy gap observed in the tetragonal structure of KCuF${}_3$ with $s=1/2$ \cite{lake2005}. We hope that more experimental results for the energy spectra at other wavevectors will be available for comparison.

In this article, we extend our microscopic approach to study the longitudinal modes in quantum antiferromagnets on a triangular lattice where the magnetic order is noncollinear hence there are possible more than one longitudinal modes. We employ the approximate SWT ground state in our calculations and find that one of the longitudinal modes is gapless for the isotropic models in the thermodynamic limit, due to the slow divergence of the structure factor, but any finite size effect or small anisotropy will induce a large energy gap at the magnetic wavevectors when compared with the counterpart of the transverse spin-wave spectra, similar to the collinear square lattice model. We organize this article as follows. For completeness, we briefly outline the main results of spin-wave theory for the triangular lattice model in Sec.~2, using the one-boson approach. We then apply our microscopic theory for the longitudinal excitations in Sec.~3, using the approximated ground state from SWT. We find that a large energy gap can be induced by a very tiny anisotropy, similar to the square lattice model. We also discuss the possible other longitudinal mode of the triangular lattice and the possible extension of our calculations to the more realistic models in the quasi-1d systems such as those hexagonal $ABX_3$-type antiferromagnets mentioned above in the last section.

\section{Spin-wave theory for triangular lattice models}

The classical ground state of the antiferromagnetic Heisenberg model on a triangular lattice consists of three alternating sublattices with spins on each sublattice align at an angle of $120^\circ$ to the other two sublattices. The distance between the two nearest-neighbor spins of the same sublattice is $\sqrt3$ of the lattice spacing which is taken as unity in this article. The spin-$s$ Heisenberg Hamiltonian is given by
\begin{equation}\label{1}
    H=J\sum_{\langle i,j\rangle}\mathbf{S}_i\cdot \mathbf{S}_j,
\end{equation}
where $J(>0)$ is the coupling parameter and the sum on $\langle i,j\rangle$ runs over all the nearest-neighbor pairs of the triangular lattice once. Following Singh and Huse \cite{Singh1} and Miyake \cite{Miyake1992} , it is convenient to transform the Hamiltonian of Eq.~\eqref{1} by rotating the quantum projection axis of the spins along the classical direction in the $xz$-plane at the six $i$-sublattices surrounding the one $j$-sublattice. This transformation leads to the following rotated spin operators,
\begin{equation}\label{2}
\begin{split}
S_i^x&\rightarrow S_i^x\cos(\mathbf{Q}\cdot \mathbf{r}_i)+S_i^z\sin(\mathbf{Q}\cdot \mathbf{r}_i),\\
S_i^y&\rightarrow S_i^y,\\
S_i^z&\rightarrow S_i^z\cos(\mathbf{Q}\cdot \mathbf{r}_i)-S_i^x\sin(\mathbf{Q}\cdot\mathbf{r}_i),
\end{split}
\end{equation}
for all $i$ sites, where $\mathbf Q=(4\pi/3,0)$ is the magnetic ordering wavevector at the corner of the hexagonal Brillouin zone of the triangular lattice (see Fig.~1a). After the rotation of Eq.~\eqref{2}, the spin-wave theory can be formulated in terms of only one set of bosons, rather than three sets originally employed \cite{Jolicoeur}. The Hamiltonian operator of Eq.~\eqref{1} after this transformation is given by
\begin{align}\label{3}
    H&=J\sum\limits_{\langle i,j\rangle}\big[ \cos[\mathbf Q\cdot(\mathbf r_i-\mathbf r_j)](S_i^xS_j^x+S_i^zS_j^z)+\Delta S_i^yS_j^y\nonumber\\
    &+\sin[\mathbf Q\cdot(\mathbf r_i-\mathbf r_j)]
    (S_i^zS_j^x-S_i^xS_j^z)\big],
\end{align}
where we have also introduced an anisotropy parameter $\Delta (\le1)$ along the $y$-axis. We will see later that even a very tiny anisotropy, such as $\Delta=1-1.5\times10^{-4}$, will induce a large energy gap for the longitudinal excitation spectrum at the magnetic wavevectors $\pm\mathbf Q$ when compared with the counterpart of the spin-wave spectra. Using the conventional Holstein-Primakoff transformations,the Hamiltonian of Eq.~\eqref{3} can be expressed in terms of boson operators $a_i$ and $a^\dagger_i$ in a series in power of $s$. The classical result for the ground-state energy is given by the ${\cal O}(s^2)$ term. The linear terms (i.e., linear in $a_i$ and $a_i^\dagger$) cancel each other out. The quadratic terms are in ${\cal O}(s)$ and are retained in the linear SWT. The cubic terms (in ${\cal O}(\sqrt s)$) and the quartic terms (in ${\cal O}(s^0)$) have been treated as perturbations for the higher-order corrections to the linear SWT \cite{Miyake1992,Chubukov1994,Chernyshev2009a}. Therefore, the Hamiltonian of the linear SWT is given by, keeping only the ${\cal O}(s^2)$ and ${\cal O}(s)$ terms and after a Fourier transofrmation for the boson operators with the Fourier component operators $a_q$ and $a^\dagger_q$,
\begin{equation}\label{4}
    H'=-\frac{3}{2}JNs^2+3Js\sum_q\left[A_qa_q^\dagger a_{-q}-\frac{1}{2}B_q(a_q^\dagger a_{-q}^\dagger+a_{q}a_{-q})\right],
\end{equation}
where $A_q$ and $B_q$ are defined by
\begin{equation}\label{5a}
    A_q=1+\left(\Delta-\frac{1}{2}\right)\gamma_q,\quad
    B_q=\left(\Delta+\frac{1}{2}\right)\gamma_q,
\end{equation}
respectively and $\gamma_q$ is defined as usual by
\begin{equation}\label{6}
    \gamma_q=\frac{1}{z}\sum_\rho e^{i\mathbf {q\cdot r}_\rho}=\frac{1}{3}\Big(\cos q_x+2\cos
    \frac{q_x}{2}\cos\frac{\sqrt{3}}{2}q_y\Big),
\end{equation}
with the summation over the nearest-neighbor index $\rho$ and the coordination number $z=6$ for the triangular lattice. The Hamiltonian $H'$ is diagonalized by the canonical Bogoliubov transformation, $a_q=u_q\alpha_q+v_q\alpha_{-q}^\dagger$,
\begin{equation}\label{8}
   H'=-\frac{3}{2}JNs(s+1)+\sum_q{\cal E}_q(\alpha_q^\dagger \alpha_q+\frac{1}{2}),
\end{equation}
where, ${\cal E}_q=zJs\,\omega_q/2$, is the spin-wave excitation spectrum with the dimensionless spectrum $\omega_q$ given by
\begin{equation}\label{9}
\omega_q=\sqrt{A_q^2-B_q^2}=\sqrt{(1-\gamma_q)(1+2\Delta\gamma_q)}\,.
\end{equation}
We plot the first Brillouin zone (BZ) of the triangular lattice and the dimensionless energy spectrum $\omega_q$ of Eq.~\eqref{9} in Fig.~{1}. This spin-wave spectrum has three zero modes, one at the center of the zone $\Gamma (\mathbf q=0$), and the other two at the corners of the BZ, the $\pm\mathbf Q$ ordering wavevectors. The magnitude of the spin-wave velocity near these points are different, with values of $v_\Gamma =3Js\sqrt3 /2$ and  $v_{\pm\mathbf Q} =3Js\sqrt3 /2\sqrt2$ respectively.
\begin{figure}[h]
\begin{center}
\subfigure[]{
   \includegraphics[scale=0.35]{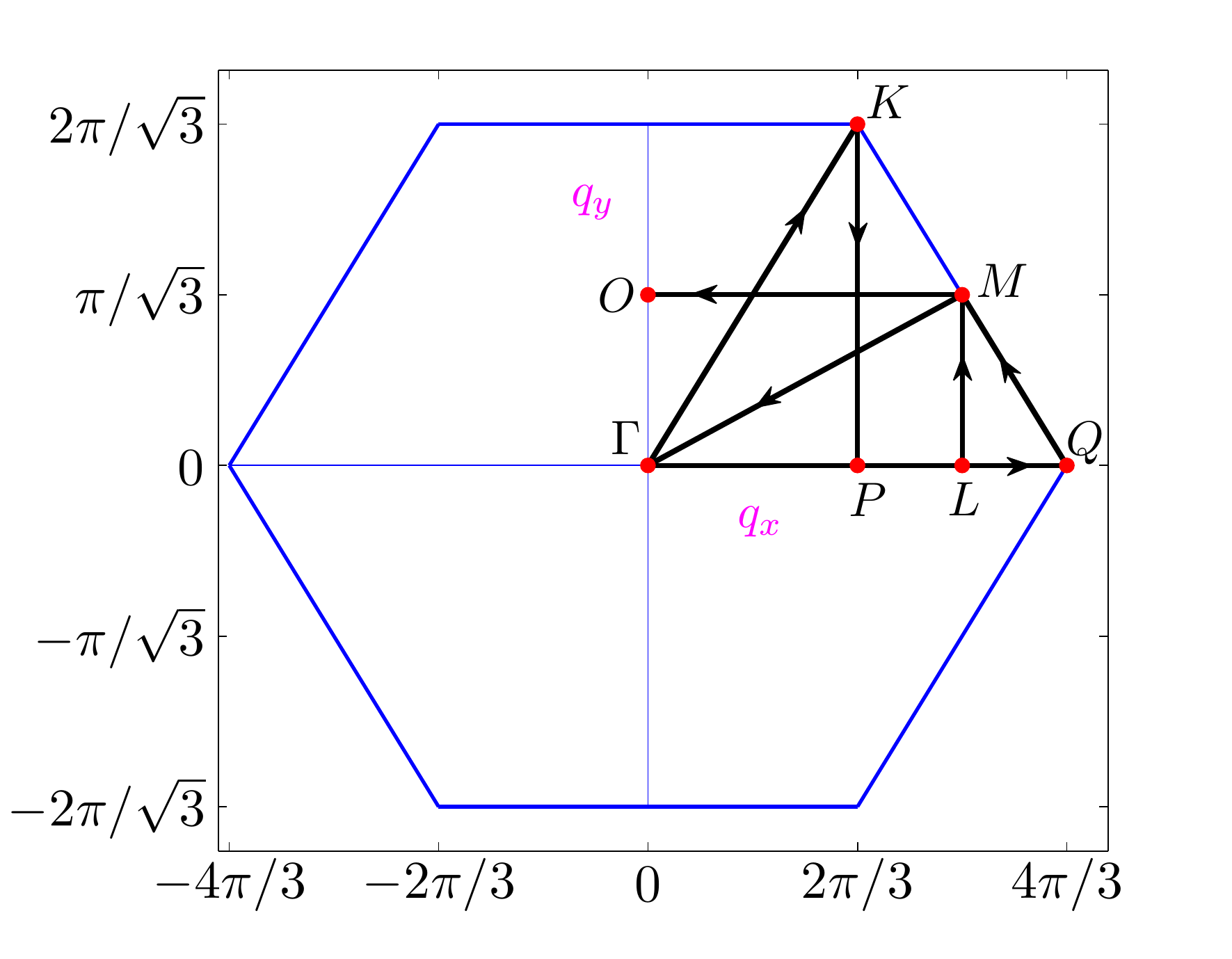}
 }
 \hspace{0.0cm}
 \subfigure[]{

   \includegraphics[scale=0.35]{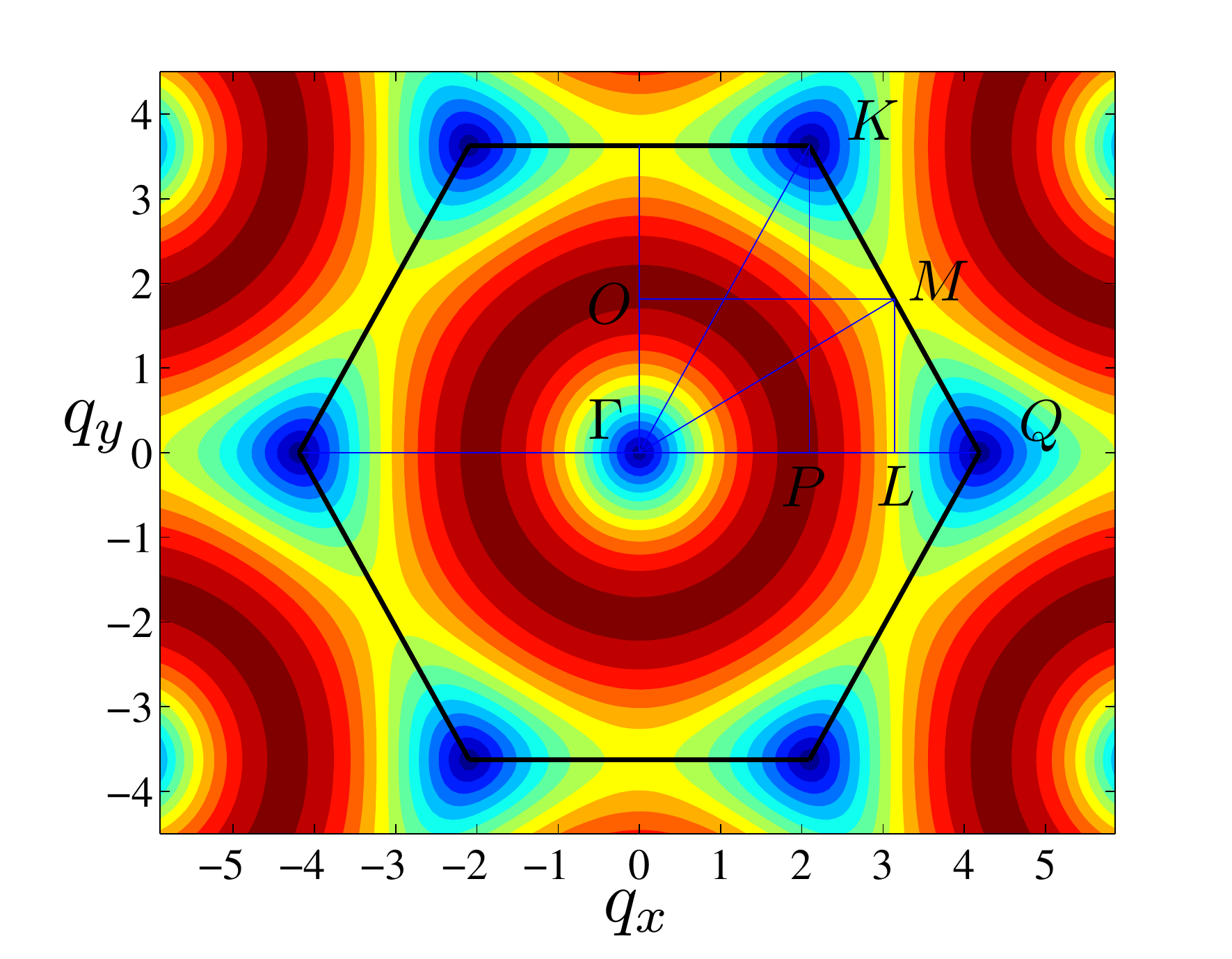}
}
\caption{\small (a) The hexagonal first Brillouin zone of a triangular lattice in reciprocal space. The coordinates of the labeled points are, $\Gamma=(0,0)$, $P=(2\pi/3,0)$, $L=(\pi,0)$, $Q=(4\pi/3,0)$, $M=(\pi,\pi/\sqrt3)$, $K=(2\pi/3,2\pi/\sqrt3)$ and $O=(0,\pi/\sqrt3)$. (b) The intensity of the linear spin-wave energy spectrum (dimensionless) $\omega_q$ of Eq.~\eqref{9} for a triangular-lattice Heisenberg model at $\Delta=1$.}
\end{center}
\end{figure}

The sublattice magnetization $m$ is given by $m= s-\eta$ with the magnon density $\eta$ defined as the ground-state expectation value of boson number operator $\eta=\langle a^\dagger_ia_i\rangle_g$ with value $\eta=0.261$ for the isotropic model in the linear SWT. This magnon density of $\eta=0.261$ per lattice site is larger than the corresponding number of $0.197$ for the square lattice model, indicating the larger quantum correction in the triangular lattice model due to the frustration. Furthermore, we notice that the three sublattice magnetization are noncollinear with $120^\circ$ between their directions, hence we may have two longitudinal modes  in the triangular lattice model, one corresponding to the $z$-component magnon-density fluctuations in our rotated spin basis discussed above, the other corresponding to the perpendicular $x$-component magnon-density fluctuations. In the following section, we will discuss the longitudinal excitations in the $z$-component magnon-density fluctuations, similar to the collinear square lattice case where there is only one longitudinal mode and leave the discussion of the other possible longitudinal mode to the last section.

\section{The longitudinal mode of the $z$-component magnon-density fluctuations in triangular lattice model}

As mentioned earlier, the longitudinal excitations in an antiferromagnetic system with a N\'eel-like long-ranged order correspond to the fluctuations in the order parameter. We have proposed a microscopic approach based on the magnon-density waves and have applied to bipartite antiferromagnetic systems where the magnetic order is collinear \cite{yang2011}. Here we extend our analysis to triangular lattice models using the one-boson approach for the approximation of the ground state and investigate the longitudinal mode of the $z$-component magnon-density fluctuations. Briefly, the excitation state of the $z$-component magnon-density waves is constructed by applying the magnon-density operator $X_q$ to the ground state $|\Psi_g\rangle$,
\begin{equation}
|\Psi_e\rangle =X_q|\Psi_g\rangle,
\end{equation}
where the density operator $X_q$, instead of the spin-raising and lowering operators $s^\pm$ for the transverse spin-wave excitations (i.e. magnons), is given by $s^z$ operator as
\begin{equation}\label{12}
X_q = \frac1N\sum_l e^{i\mathbf q\cdot r_l} s^z_l,\quad q>0,
\end{equation}
where the index $l$ runs over all lattice sites and the condition $q>0$ ensures that the excited state $|\Psi_e\rangle$ is orthogonal to the ground state. We notice that in our rotated spin basis of Eq.~(2), the three sublattice magnetizations all point in the same $z$-direction. The longitudinal excitation spectrum is then given by
\begin{equation}\label{13}
    E(q)=\frac{N(q)}{S(q)},
\end{equation}
where $N(q)$ is given by the ground-state expectation value of a double commutator as
\begin{equation}\label{14}
    N(q)=\frac{1}{2}\langle[X_{-q},[H,X_q]]\rangle_g,
\end{equation}
and the state normalization integral $S(q)$ is the structure factor of the lattice model
\begin{equation}\label{15}
    S(q)=\langle X_{-q}X_q\rangle_g=\frac{1}{N}\sum_{l,l'}e^{i\mathbf q.(\mathbf r_{l}-\mathbf r_{l'})}\langle s_l^zs_{l'}^z\rangle_g.
\end{equation}
In Eqs.~\eqref{14} and \eqref{15}, the notation $\langle\dots\rangle_g$ indicates the ground-state expectation. It is straightforward to derive the double commutator for triangular lattice by employing the one-sublattice Hamiltonian of Eq.~\eqref{3} after the rotation to obtain the following general expression,
\begin{align}\label{16}
   N(q)&=\frac{1}{4N}J\sum_{l,\rho}\Big[\frac{1}{2}(1+2\Delta)
   (1+\gamma_q)\langle S_l^+S_{l+\rho}^+\rangle_g
   +\frac{1}{2}(1-2\Delta)(1-\gamma_q)\langle S_l^+S_{l+\rho}^-\rangle_g
   \nonumber\\
   &-\sin[\mathbf Q\cdot(\mathbf r_l-\mathbf r_{l+\rho})]
    \big[\langle S_l^zS_{l+\rho}^x\rangle_g-\langle S_l^xS_{l+\rho}^z\rangle_g\big]\Big].
\end{align}
Now we apply the approximation of the linear SWT for the ground state in the expectations $\langle\dots\rangle_g$ and obtain
\begin{equation}\label{17}
   N(q)=\frac{1}{4}sJ\sum_\rho\Big[(1+2\Delta)
   (1+\gamma_q)\tilde g_\rho+(1-2\Delta)
   (1-\gamma_q)\tilde g'_\rho\Big],
   \end{equation}
where the transverse correlation functions $\tilde g_r$ and $\tilde g'_r$ are defined as
\begin{equation}\label{18}
    \tilde g_r=\frac{1}{2s}\langle S_l^+S_{l+r}^+\rangle_g,\quad   \tilde g'_r=\frac{1}{2s}\langle S_l^+S_{l+r}^-\rangle_g,
\end{equation}
both independence of index $l$ due to the lattice translational symmetry.
Their Fourier transformations are obtained as, using the approximation
of the linear SWT for the ground state in the one-boson approach discussed in Sec.~2,
\begin{equation}\label{20}
    \tilde g_q=\frac{1}{2}\frac{B_q}{\sqrt{A_q^2-B_q^2}},\quad \tilde g'_q=\frac{1}{2}(\frac{A_q}{\sqrt{A_q^2-B_q^2}}-1),
\end{equation}
and where $A_q$ and $B_q$ are as given by Eqs.~\eqref{5a}. We obtain the numerical results at the isotropic point $\Delta=1$ as $\tilde g_\rho=0.258$ and $\tilde g_\rho'=0.034$ for all the six nearest neighbors. As can be seen, $N(q)$ is dominated by $\tilde g_\rho$.

Within the same approximation of the linear SWT for the ground state, the structure factor $S(q)$ of Eq.~\eqref{15} is obtained as
\begin{equation}\label{22}
    S(q)=\eta+\frac{1}{N}\sum_{q'}\eta_{q'}\eta_{q+q'}+\frac{1}{N}
    \sum_{q'}\tilde g_{q'}\tilde g_{q+q'},
\end{equation}
where $\tilde g_q$ is as given by Eq.~\eqref{20}, $\eta$ is the magnon density as discussed before and
\begin{equation}\label{24}
\eta_q=\tilde g'_q=\frac{1}{2}(\frac{A_q}{\sqrt{A_q^2-B_q^2}}-1).
\end{equation}
We notice that the structure factor of Eq.~\eqref{22} involves the double magnon spectrum function $\omega_q\omega_{q+q'}$ in the integrals, indicating some effects of the interactions between magnons has been included. We then obtain the longitudinal excitation spectrum $E(q)$ of Eq.~\eqref{13} using the approximations of Eqs.~\eqref{17} and \eqref{22}. By numerical evaluation, we notice that this spectrum of the longitudinal mode is the gapless in the thermodynamic limit, approaching zero $E(q)\to 0$ at both $ q\to0$ and ${\mathbf q}\to\pm \mathbf Q$. Detailed numerical analysis shows that the gapless spectrum is due to the slow, logarithmic divergence in both the second and third terms in the structure factor $S(q)$ of Eq.~\eqref{22}. More specifically, near $\Gamma$ and $\pm\mathbf Q$, we find that $S(q)\propto -\ln q$, and thus the excitation spectrum $E(q)\propto-1/\ln q$ as $q\to0$ or $q\to\pm\mathbf Q$, but with different coefficients for the $\Gamma$ point from that of the $\pm\mathbf Q$ points. This later feature is reminiscent of the different spin-wave velocities at these points as discussed in the previous section. The major difference between the longitudinal and transverse modes near $\Gamma$ and $\pm\mathbf Q$ points will become apparent once we introduce the anisotropy as discussed in the followings.

The logarithmic behaviors of the structure factor and of the energy spectrum of triangular lattice model is similar to that of the square lattice model studied earlier \cite{Xian2006,Xian2007,yang2011}, where we state that the spectrum is quasi-gapped as any finite finite size effect or anisotropy will induce a large energy gap when compared with the counterparts of spin-wave spectrum. Here we consider the effect of anisotropy in the $y$-axis as given by Eq.~\eqref{3}. For a small value of $\Delta=1-1.5\times10^{-4}$, we obtain energy gap value of $0.203zsJ$ at the ordering wavevector $\pm\mathbf Q$, compared with the much smaller gap value of $0.0075zsJ$ of the corresponding spin-wave spectrum. More specifically, we find that the longitudinal energy gap value is proportional to $1/[-\ln(1-\Delta)]$, in contrast to the spin-wave gap which is proportional to $\sqrt{1-\Delta}$, when $\Delta\rightarrow1$. We notice that the spectrum is still gapless at $\Gamma$ due to the fact that the anisotropy $\Delta$ is introduced in the $y$-component in the Hamiltonian of Eq.~\eqref{3} while the spins are ordered along the rotated $z$-axis. In order to make further comparison between the longitudinal mode and the transverse spin-waves mode, we plot both the spectra with $\Delta=1-1.5\times10^{-4}$ in Fig.~\ref{6} along the path $(LM\Gamma KPQMO)$ of the BZ. The different gap values for the longitudinal and transverse mode at both $K$ and $Q$ points can be clearly seen. Both the longitudinal and spin-wave spectra at $\Gamma(\mathbf q=0)$ are still gapless where $\gamma_{\mathbf q}=1$ for any value of $\Delta (\le1)$. Furthermore, there are slight different peak values at for the longitudinal spectrum but for the spin-wave spectrum the peak values are nearly equal. This is again similar to the case of the square lattice model discussed earlier \cite{yang2011}, indicating some effects of the interaction between magnons in the longitudinal mode.
\begin{figure}[h]
\centering
\includegraphics[scale=0.5]{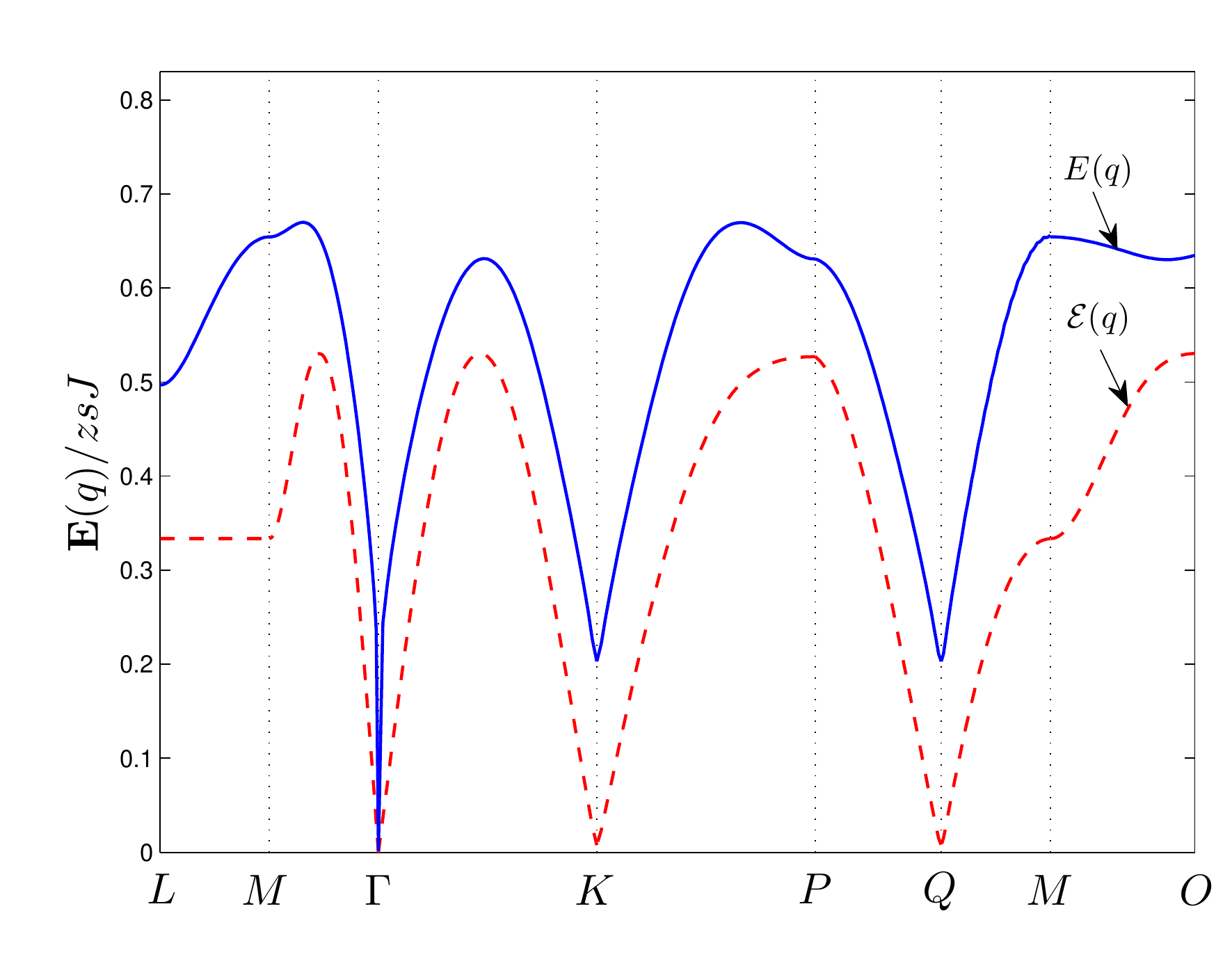}\\
  \caption{The excitation spectrum $E(q)$ of the longitudinal mode together with spin-wave excitation spectrum ${\cal E}(q)$ along ($LM\Gamma KPQMO$) of the BZ with an anisotropy $\Delta=1-1.5\times10^{-4}$. The gap value at $K$ and $Q$ points is $0.203zsJ$ for the longitudinal mode, and $0.0075zsJ$ for the transverse spin-wave modes.}
  \label{6}
\end{figure}

We now turn our attention to the difference in the excitation spectra between the triangular and square lattice models. This can be easily demonstrated by the linear SWT for both models. Due to the frustrations in the triangular lattice, there are three sublattice magnetizations with $120^\circ$ different orientation to one another rather than the two sublattices with complete opposite orientation ($180^\circ$) in square lattice. The anisotropy parameter $\Delta\le1$ is introduced in the $y$ components of the Hamiltonian of Eq.~\eqref{3} rather than in the usual the ordered direction of the $z$ components for the square lattice model, hence the spin-wave spectra of the triangular anisotropic model are still gapless at the $\Gamma$ point. Furthermore, the anisotropy parameter $\Delta$ inside the square root formula of the spin-wave spectrum of Eq.~\eqref{9} is linear rather than the usual quadratic form for the bipartite models such as the square lattice model. All these features of the triangular lattice model contribute to the slightly softer nature of the longitudinal model than that of the square lattice model. In particular, the energy gap values for the longitudinal mode is about $0.203zsJ$ at $Q$ and $0$ at $\Gamma$ for the triangular lattice model with the anisotropy $\Delta=1-1.5\times10^{-4}$ as given before, comparing with the bigger energy gap values of $0.76zsJ$ at $\mathbf Q$ and $0.44zsJ$ at $\Gamma$ for the square lattice model at the similar value of the anisotropy $\Delta=1+1.5\times10^{-4}$ which enters the Hamiltonian in the ordered $z$ component \cite{yang2011}. More significantly, due to the fact that antiferromagnetic order is noncollinear in triangular lattice, in additional to the above discussed longitudinal model corresponding to the $z$-component order parameter fluctuations, we may have another longitudinal mode corresponding to the perpendicular $x$-axis fluctuations. We leave further discussion to the following section.

\section{Summary and discussion}

In this paper, we have applied our microscopic analysis based on the magnon-density waves to study the longitudinal excitations of the $z$-component order parameter fluctuations for a triangular lattice antiferromagnets, with the results similar to those of the square lattice model. In particular, the longitudinal spectrum for the isotropic model has zero modes at the zone center and the magnetic wavevectors due to the slow, logarithmic divergence of the structure factor. Also similar to the square lattice model, any finite size effect or anisotropy will induce large energy gaps at the magnetic wavevectors when compared with the counterparts of the transverse spin-wave excitations. Furthermore, we find the longitudinal modes of the triangular lattice models are in general softer than that of the square lattice model due to the frustrations in the triangular lattice model. Due to the fact that the magnetic order in the triangular lattice is noncollinear, there may exist another longitudinal mode, corresponding to the order parameter fluctuations in the perpendicular $x$ direction. Instead of the density operator of Eq.~\eqref{12}, we use the operator $X_q=\frac{1}{N}\sum_le^{i{\bf q}.r_l}s_l^x$, for the density fluctuations in the x-direction. We will report the results for this longitudinal mode elsewhere.

One remaining question is the intrinsic lifetime of the longitudinal modes for triangular lattice model. As the results in Sec.~3 demonstrate that the energy of the longitudinal mode is always higher than that of the transverse spin-wave excitations, we expect the longitudinal mode in the pure triangular lattice model may be unstable against decaying into two or more magnons, as emphasized in Ref.~28 and in a more recent discussion \cite{kulik2011}. One way to find the stable longitudinal modes is to consider some frustrated models in triangular lattices near a quantum critical point where the sublattice magnetization is much reduced \cite{PhysRevB.79.174405,Kurosaki}. A more realistic way is to consider some quasi-1d hexagonal systems as mentioned in Sec.~I. It is therefore particularly interesting to extend our present analysis to the hexagonal $ABX_3$-type antiferromagnets with both spin quantum number $s=1$ (CsNiCl${}_3$ and RbNiCl${}_3$)  \cite{Steiner1987,Tun1990} and $s=5/2$ (CsMnI${}_3$) \cite{PhysRevB.43.679,Kenzelmann2002}, where the longitudinal energy gaps were first observed at the magnetic wavevectors. The spin lattice structure on the basal planes of these systems are triangular as that discussed here. The field theory approach proposed by Afflect based on Haldane's theory of spin-1 (or integer spin quantum number) chain with five fitting parameters is able to explain many features observed, but there are still some disagreements particularly for the data away from the minimum energy gap at the antiferromagnetic wavevectors \cite{Kenzelmann2002}. We believe that the longitudinal excitations of the quasi-1d antiferromagnets is more general than the extension of Haldane's theory of integer spin quantum chains and that our microscopic analysis based on the magnon-density waves as presented here may provide a more general description of such longitudinal modes without any fitting parameter other than those in the model Hamiltonian.

%%%%%%%%%%%%%%%%%%%%%%%%%%%%%%%%%%%%%%%%%%%%%%%%%%%%%%%%%%%%%%%%%%%%%%
\bibliographystyle{h-physrev3}

\end{document}